\documentclass{article}

\usepackage[breaklinks,hidelinks]{hyperref}

\PassOptionsToPackage{numbers, compress}{natbib}
\usepackage[final]{nips_2017}

\usepackage[utf8]{inputenc} 
\usepackage[T1]{fontenc}    
\usepackage{hyperref}       
\usepackage{url}            
\usepackage{booktabs}       
\usepackage{amsfonts}       
\usepackage{nicefrac}       
\usepackage{microtype}      
\usepackage{graphicx}
\usepackage{wrapfig}
\usepackage{multirow}
\usepackage[font={small}]{caption}

\usepackage{floatrow}
\newfloatcommand{capbtabbox}{table}[][\FBwidth]

\usepackage{blindtext}

\title{Vehicle Communication Strategies \\
for Simulated Highway Driving}

\author{
  Cinjon Resnick\\
  NYU\\
  \texttt{cinjon@nyu.edu} \\
  \And
  Ilya Kulikov\\
  NYU\\
  \texttt{kulikov@cs.nyu.edu}\\
  \And
  Kyunghyun Cho\\
  NYU, FAIR\\
  \texttt{kyunghyun.cho@nyu.edu}\\
  \And
  Jason Weston\\
  FAIR, NYU\\
  \texttt{jase@fb.com}
}

\begin{document}

\maketitle

\begin{abstract}
Interest in emergent communication has recently surged in Machine Learning. The focus of this interest has largely been either on investigating the properties of the learned protocol or on utilizing emergent communication to better solve problems that already have a viable solution. Here, we consider self-driving cars coordinating with each other and focus on how communication influences the agents' collective behavior. Our main result is that communication helps (most) with adverse conditions.
\end{abstract}

\section{Introduction}

Car accidents are recognized as a serious problem. Although modern vehicles are often equipped with many accident-avoidance systems, the number of highway-related fatalities in the US alone is approximately 32,000 a year. As a response to this issue, the US Department of Transportation (DOT) has issued a notice~\cite{notice} of proposed rule-making (NPRM) requiring the installation of vehicle-to-vehicle (V2V) communication capabilities in all new cars by 2023. This is expected to become a federal motor vehicle safety standard.

The approach is that vehicles will issue messages alerting each other of potential safety concerns so as to act upon these messages and avoid accidents. The NPRM \cite{protocol} enumerates the causes of vehicle crashes and the information that will a will aid in avoiding them, yet there are many uncertainties in the proposal and requests for comments from stakeholders. For instance, the proposal expresses its uncertainty in which type of information must be sent as part of a safety message:
\begin{quote}
\it
We tentatively believe that speed, heading, acceleration, and yaw are the most relevant pieces of information about a vehicle’s moment. Essentially, we propose to measure the rate at which the sending device’s location is changing and also any changes to that rate at which a device’s location is changing...
\end{quote}
In another case, the proposal requests for comments on the specification of transmission range:
\begin{quote}
\it We ask for comment on [the minimum V2V range limit]. Is there any reason that the agency should require a maximum transmission range as well as a minimum? Should the agency choose a different minimum range requirement? What would be appropriate alternative minimum and maximum transmission range values?
\end{quote}

This uncertainty calls for a scientific investigation. In this work, we explore the possibility of studying the effect of communication in vehicle coordination in the context of reinforcement learning. In doing so, we create our own simplified simulation in which we can deploy multiple vehicles, each with a distinct goal. We then train these vehicles and demonstrate that communication, specifically an engineered protocol used as input to the policy, greatly improves both safety and efficiency in adverse conditions. We also investigate emergent communication approaches but our efforts there are still preliminary.

\subsection{Related Work}

\begin{wrapfigure}{R}{0.3\textwidth}
  \vspace{-16mm}
\centering
  \includegraphics[width=\columnwidth]{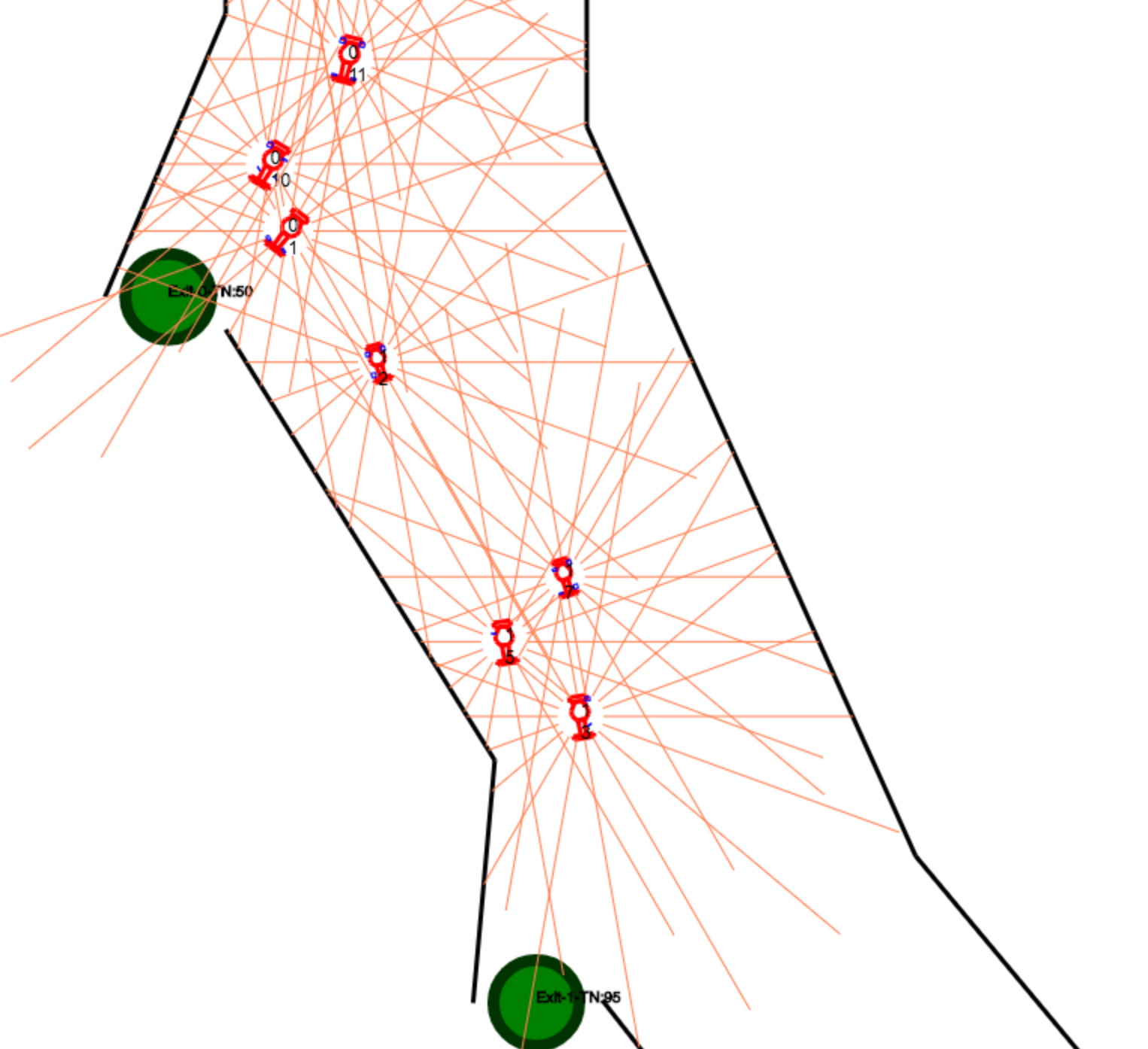}
  
  \caption{Rendering after the cars have separated.}
  \label{fig:env}
  
\end{wrapfigure}

\paragraph{Communication in multi-agent scenarios}

During the past couple of years, there has been a surge of publications in the area of multi-agent communication. At a high level, we can categorize them into two families. In the first family, communication is restricted to natural language, enabling us to study the effect and properties of natural languages in a setting richer than simple supervised learning. These include communication based machine translation~\citep{DBLP:journals/corr/abs-1710-06922}, learning-to-negotiate~\citep{lewis2017deal} and visually-grounded dialog agents~\citep{das2017learning}. Another angle has focused more on solving a target task by allowing multiple agents to develop their own communication protocol. These include simple traffic navigation~\citep{NIPS2016_6398}, cooperative riddle solving~\citep{foerster2016learning,evtimova2017emergent} and multi-agent reinforcement learning~\citep{mordatch2017emergence,lowe2017multi}. 

Our work sits in between these two directions of research in that we study the multi-agent coordination problem augmented with either a manually-designed communication protocol or an emergent one, and compare them against a no-communication setting. Additionally, our research is aimed towards solving a problem that today has no good solution.

\paragraph{Autonomous driving}

Fully learning-based autonomous driving has received renewed interest since the early work by \citet{muller2006off}. There is related work on training a single vehicle to drive~\citep[see, e.g.,][and references therein]{bojarski2016end,zhang2016query,chowdhuri2017multi}. There is also work trying to train multiple vehicles collectively to better coordinate amongst themselves~\citep[see, e.g.,][]{shalev2016safe}, albeit without communication. Our work falls into the second category, however, with the novelty that we allow vehicles to directly message each other, potentially allowing them to develop a more efficient coordination strategy.

\begin{wraptable}{R}{0.35\textwidth}
\centering
{
\small

\begin{tabular}{ | l r |}
 1-2 & Exit Position \\
 3 & Max Speed \\
 4-5 & Velocity (X, Y) \\
 6-7 & Current Position \\
 8 & Steering Wheel Angle \\
 9 & Size \\
 10 & Acceleration \\
 11 & Car Angle \\
 12 & Time \\
 13-22 & Path History \\
 23-40 & Lidar Distances
\end{tabular}
}
\caption{Personal observation.}
\label{table:personalobs}
\end{wraptable}

\section{Simulation Setup}

Our simulator involves multiple vehicles, driven by a single shared policy, each tasked with driving as fast as possible to a highway exit without crashing into each other

We use Box2D~\citep{box2d}, a Python gaming framework, to build the highway and use a car implementation from OpenAI Gym~\citep{gym}. Every episode consists of twelve cars. The vehicle's size, exit, and starting position are selected randomly from five, two, and fifteen possibilities respectively. The length and angle of the highway also change randomly at each episode, and there are no lane markings, which makes driving even more challenging. These choices ensure diversity in the driving scenarios. See Fig.~\ref{fig:env} for a rendering of this environment.

Our reward structure is meant to encourage the agents to quickly drive to the exit without crashing. Consequently, we settled on $+60$ for a successful exit, $-60$ for a crash or going past the exit. A negative reward of $-0.5$ is given each step.

\begin{wraptable}{R}{0.35\textwidth}
\centering
{
\small
\begin{tabular}{ | l r | }
 1 & Car Id \\
 2 & Global Message Id \\
 3 & Episode time step \\
 4 & Current X Position \\
 5 & Current Y Position \\
 6 & Car Speed \\
 7 & Car Angle \\
 8 & Car Acceleration \\
 9-18 & Path History \\
 19 & Hard Brake Indicator \\
 20 & Steering wheel angle \\
 21 & Car Size
\end{tabular}
}
\caption{V2V message description.}
\label{table:v2vobs}
\end{wraptable}

\section{Policy Setup}

\paragraph{Observation and action}

We share a single policy across all the vehicles in the highway. In our baseline model without communication, each agent receives a 40-dimensional observation vector every time step as described in Table~\ref{table:personalobs}. When communication is allowed, the observation vector is augmented with the messages from all other nearby vehicles. In the case of the V2V message protocol, there are 21 additional dimensions, delineated in Table~\ref{table:v2vobs}. Note that we do not adhere to the full specification of V2V \citep{protocol} as some fields suggested in the protocol do not fit the proposed environment. Given such an observation, the policy outputs a tuple of three continuous actions: acceleration, steering wheel angle, and brake.

We do not impose any restriction other than the structures of the observations and actions as described above. This gives the policy full freedom in developing novel strategies based on its observations as well as messages from other vehicles. In the case of emergent communication especially, we expect this freedom to allow the policy to develop a more effective and efficient protocol than which can obtained using an engineered approach.

\paragraph{Policy architecture}

We use a $\tanh$ nonlinear feed-forward neural network with two hidden layers (200 and 100 units, respectively) and column norm initialization for each of the policy and value networks. Together, they output the mean of a Gaussian distribution per action. 

There are a number of ways to incorporate the messages into the model. From the preliminary experiment, it was found most effective to concatenate all the messages together, non-linearly project them into a fixed size 32-dimensional input, and then concatenate that result with the personal observations as the input to the policy. 

\paragraph{Learning}

After each episode, we accumulate a sequence of state-action-reward tuples per vehicle. We consider each such vehicle-specific episode as an independent episode when training a policy. We use a recently proposed proximal policy algorithm~\citep[PPO,][]{ppo} based on the TensorFlow implementation by \citet{DBLP:journals/corr/abs-1709-02878}.

We found that randomly dropping the messages $10$\% of the time improved performance.
We implemented two variants. The first variant works per agent by dropping a received message randomly. The second variant, on the other hand, randomly drops all the messages across all the vehicles at each time step. In this work, we use the first variant.

\section{Analysis}

\paragraph{Evaluation Metrics}

One target scenario for a car is to get to its destination as fast as possible while never crashing. Consequently, our evaluation metrics were designed around safety and speed objectives. For both adverse and regular conditions, we measured how often the cars succeeded as opposed to crashing or passing their exits, how often \textit{all} of the cars finished the episode, and the mean number of steps for the failed cars and the successful cars. The adverse condition we consider is that of fog decreasing lidar range uniformly. 

We report results as the average of three evaluation runs with different seeds. While training was unstable - a priority to address in future research - evaluation runs across a single model were consistent.

The experimental setup we have described does not demand that the cars do perfectly on safety, but that they do a reasonable job such that marked improvements over a baseline without communication could be considered valid. Thus, we tuned the difficulty of our environment such that our best baseline could only achieve approximately $90\%$ success rate per car.

\paragraph{Results}

\begin{wraptable}{R}{0.7\textwidth}
\centering
\caption{Models trained without adverse (foggy) conditions.}
\label{table:exadverse}
{
\small

\begin{tabular}{|l|c|c|c|c|c|}
\hline
\multicolumn{1}{|c|}{\multirow{2}{*}{Model}} & \multicolumn{1}{c|}{\multirow{2}{*}{\begin{tabular}[c]{@{}c@{}}Eval \\ Conditions\end{tabular}}} & \multicolumn{1}{c|}{\multirow{2}{*}{\begin{tabular}[c]{@{}c@{}}Flat \\ success\end{tabular}}} & \multicolumn{1}{c|}{\multirow{2}{*}{\begin{tabular}[c]{@{}c@{}}Episode \\ Success\end{tabular}}} & \multicolumn{1}{c|}{\multirow{2}{*}{\begin{tabular}[c]{@{}c@{}}Mean Length \\ Success\end{tabular}}} & \multicolumn{1}{c|}{\multirow{2}{*}{\begin{tabular}[c]{@{}c@{}}Mean Length\\ Fail\end{tabular}}} \\
\multicolumn{1}{|c|}{} & \multicolumn{1}{c|}{} & \multicolumn{1}{c|}{} & \multicolumn{1}{c|}{} & \multicolumn{1}{c|}{} & \multicolumn{1}{c|}{} \\ \hline
Baseline & Sunny & 0.9302 & 0.6348 & 275.7 & 273.6 \\ \hline V2V & Sunny & 0.9385 & 0.6304 & 308.8 & 307.1 \\ \hline
Baseline & Foggy & 0.8302 & 0.5456 & 278.2 & 231.7 \\ \hline
V2V & Foggy & 0.8877 & 0.6104 & 307.7 & 280.3  \\ \hline
\end{tabular}
}
\end{wraptable}

Table~\ref{table:exadverse} presents results for models trained on normal sunny conditions without fog, but with evaluation for both the sunny and foggy settings. The best baseline model succeeded at a rate of $93\%$ per car and $63.5\%$ per episode in the sunny environment. It starts to learn a sufficiently good policy after $600$ episodes. While the model took longer to learn (approximately 2,400 episodes) with the V2V communication, the per-car success rate improved at the expense of a $10\%$ speed reduction. The gap in the success rate between the baseline and the one with the V2V communication grows when evaluated in foggy conditions. These relative numbers are roughly the same, albeit both models lose some efficacy.

\begin{wraptable}{R}{0.7\textwidth}
\centering
\caption{Models trained with adverse (foggy) conditions.}
\label{table:adverse}
{
\small
\begin{tabular}{|l|c|c|c|c|c|}
\hline
\multicolumn{1}{|c|}{\multirow{2}{*}{Model}} & \multicolumn{1}{c|}{\multirow{2}{*}{\begin{tabular}[c]{@{}c@{}}Eval \\ Conditions\end{tabular}}} & \multicolumn{1}{c|}{\multirow{2}{*}{\begin{tabular}[c]{@{}c@{}}Flat \\ Success\end{tabular}}} & \multicolumn{1}{c|}{\multirow{2}{*}{\begin{tabular}[c]{@{}c@{}}Episode \\ Success\end{tabular}}} & \multicolumn{1}{c|}{\multirow{2}{*}{\begin{tabular}[c]{@{}c@{}}Mean Length \\ Success\end{tabular}}} & \multicolumn{1}{c|}{\multirow{2}{*}{\begin{tabular}[c]{@{}c@{}}Mean Length\\ Fail\end{tabular}}} \\
\multicolumn{1}{|c|}{}                       & \multicolumn{1}{c|}{}                                                                         & \multicolumn{1}{c|}{}                                                                         & \multicolumn{1}{c|}{}                                                                            & \multicolumn{1}{c|}{}                                                                                & \multicolumn{1}{c|}{} \\ \hline \hline
Baseline           & Sunny        & 0.7951 & 0.2340     & 296.6         & 281.7      \\ \hline
V2V & Sunny         & 0.9103   &   0.5257       &   320.9              &    320.6           \\ \hline
Baseline           & Foggy          & 0.7031  & 0.2568    & 291.4         & 274.6      \\ \hline
V2V & Foggy          & 0.8571 &  0.4606   &   325.7   &   314.9   \\ \hline
\end{tabular}
}
\end{wraptable}

We also tried training the model in adverse conditions. This is where we expect the communication to be most advantageous because now the model can learn to utilize the messages from other cars in situations where its lidar is ineffective. The results, shown in Table~\ref{table:adverse}, verify this and demonstrate that the V2V protocol maintains a high level of success rate while only slightly reducing the speed from the policy learned in sunny conditions.

One notable statistic in the tables is that the models trained in sunny conditions do better when evaluated in foggy conditions than the models trained in foggy conditions do. This is because we set up the evaluation process to match the training conditions where only 10\% of the episodes were foggy. Thus the sunny models do much better in the 90\% of evaluation runs which are sunny than the foggy models do on the 10\% of foggy evaluations.

\section{Discussion}
In this paper, we have explored using reinforcement learning to model a real world scenario that is otherwise very difficult to test. Our current results suggest that the proposed protocol will in fact make the road safer in adverse conditions and enable cars to go faster. There are of course caveats to this claim, a major one being that our simulation is a limited interpretation of actual driving scenarios. 

One aspect of self-driving cars that is of particular interest is that the agents could utilize the messages in ways that go beyond what an engineer would program. Whereas an engineer might program a few specific reactions to any given safety message, the size of the search space is so large that they probably will not think of the best solutions. A learned policy on the other hand can use messages in surprising ways. We have seen this recently with AlphaGo \cite{SilverHuangEtAl16nature} finding moves in Weiqi that humans players had not discovered after many centuries. In future research, we will compare the model's learned policies to fixed safety responses.

\paragraph{Preliminary result: emergent communication}

Instead of using the fixed V2V protocol, we can let the model design its own communication protocol by allowing it to synthesize a message itself. This type of emergent communication research has blossomed recently, with most prior work focusing on understanding or yielding attributes in the generated language such as compositionality \cite{DBLP:journals/corr/AndreasK17}. We are unaware of any research such as ours that utilizes this method for finding solutions to real world problems that don't already have a viable approach.

In preliminary experiments, we explored two variants of emergent communication. `Continuous' was very similar to the described V2V model. The difference was that each agent emitted a fixed number of continuous message actions, which were then fed to other agents instead of the V2V protocol. In `Select', each agent selected which of the twelve dimensions of the V2V protocol it wanted to include in a message. Their message then consisted of those parts with the rest of the protocol zeroed out.

We hypothesized that `continuous' would perform better than the V2V protocol, because the agent could include information outside of the V2V protocol or it may be encoded in a way that is easier for a receiver to decode. We hypothesized that `select' would be at least as good as the V2V protocol and would give us interpretable insight into what parts of the protocol were most important. The policies using either of these schemes were however no better and frequently converged to a degenerate solution in which the vehicles spent the entire episode turning in one place. Thus, finding ways to stabilize training and improve these models is important future research.

\subsubsection*{Acknowledgments}

We thank Tudor Achim, Ilya Kostrikov, Adam Bouhenguel, and Martin Arjovsky for helpful discussions. This work was partly supported by NVIDIA (Project: "NVIDIA - NYU Autonomous Driving Collaboration").

\bibliographystyle{abbrvnat}
\bibliography{bibliography}

\end{document}